\documentclass[aps,prl,twocolumn,superscriptaddress,showpacs,preprintnumbers,amsmath,amssymb]{revtex4}
\usepackage{graphicx} 
\usepackage{dcolumn}  
\usepackage{epstopdf}
\usepackage{amsmath}
\usepackage{verbatim}
\usepackage{xcolor}
\usepackage{hyperref}
\usepackage[percent]{overpic}
\graphicspath{{ps}}

\usepackage[running]{lineno}

\begin{document}


\title{ \quad\\[0.5cm] Observation of {\boldmath $X(3872)$} in {\boldmath $B \to X(3872) K\pi$} decays }
\noaffiliation
\affiliation{University of the Basque Country UPV/EHU, 48080 Bilbao}
\affiliation{Beihang University, Beijing 100191}
\affiliation{University of Bonn, 53115 Bonn}
\affiliation{Budker Institute of Nuclear Physics SB RAS and Novosibirsk State University, Novosibirsk 630090}
\affiliation{Faculty of Mathematics and Physics, Charles University, 121 16 Prague}
\affiliation{University of Cincinnati, Cincinnati, Ohio 45221}
\affiliation{Deutsches Elektronen--Synchrotron, 22607 Hamburg}
\affiliation{Justus-Liebig-Universit\"at Gie\ss{}en, 35392 Gie\ss{}en}
\affiliation{Gifu University, Gifu 501-1193}
\affiliation{The Graduate University for Advanced Studies, Hayama 240-0193}
\affiliation{Gyeongsang National University, Chinju 660-701}
\affiliation{Hanyang University, Seoul 133-791}
\affiliation{University of Hawaii, Honolulu, Hawaii 96822}
\affiliation{High Energy Accelerator Research Organization (KEK), Tsukuba 305-0801}
\affiliation{IKERBASQUE, Basque Foundation for Science, 48013 Bilbao}
\affiliation{Indian Institute of Technology Bhubaneswar, Satya Nagar 751007}
\affiliation{Indian Institute of Technology Guwahati, Assam 781039}
\affiliation{Indian Institute of Technology Madras, Chennai 600036}
\affiliation{Institute of High Energy Physics, Chinese Academy of Sciences, Beijing 100049}
\affiliation{Institute of High Energy Physics, Vienna 1050}
\affiliation{Institute for High Energy Physics, Protvino 142281}
\affiliation{INFN - Sezione di Torino, 10125 Torino}
\affiliation{Institute for Theoretical and Experimental Physics, Moscow 117218}
\affiliation{J. Stefan Institute, 1000 Ljubljana}
\affiliation{Kanagawa University, Yokohama 221-8686}
\affiliation{Institut f\"ur Experimentelle Kernphysik, Karlsruher Institut f\"ur Technologie, 76131 Karlsruhe}
\affiliation{Kennesaw State University, Kennesaw GA 30144}
\affiliation{Korea Institute of Science and Technology Information, Daejeon 305-806}
\affiliation{Korea University, Seoul 136-713}
\affiliation{Kyungpook National University, Daegu 702-701}
\affiliation{\'Ecole Polytechnique F\'ed\'erale de Lausanne (EPFL), Lausanne 1015}
\affiliation{Faculty of Mathematics and Physics, University of Ljubljana, 1000 Ljubljana}
\affiliation{University of Maribor, 2000 Maribor}
\affiliation{Max-Planck-Institut f\"ur Physik, 80805 M\"unchen}
\affiliation{School of Physics, University of Melbourne, Victoria 3010}
\affiliation{Moscow Physical Engineering Institute, Moscow 115409}
\affiliation{Moscow Institute of Physics and Technology, Moscow Region 141700}
\affiliation{Graduate School of Science, Nagoya University, Nagoya 464-8602}
\affiliation{Kobayashi-Maskawa Institute, Nagoya University, Nagoya 464-8602}
\affiliation{Nara Women's University, Nara 630-8506}
\affiliation{National Central University, Chung-li 32054}
\affiliation{National United University, Miao Li 36003}
\affiliation{Department of Physics, National Taiwan University, Taipei 10617}
\affiliation{H. Niewodniczanski Institute of Nuclear Physics, Krakow 31-342}
\affiliation{Nippon Dental University, Niigata 951-8580}
\affiliation{Niigata University, Niigata 950-2181}
\affiliation{Osaka City University, Osaka 558-8585}
\affiliation{Pacific Northwest National Laboratory, Richland, Washington 99352}
\affiliation{Panjab University, Chandigarh 160014}
\affiliation{Peking University, Beijing 100871}
\affiliation{University of Pittsburgh, Pittsburgh, Pennsylvania 15260}
\affiliation{Punjab Agricultural University, Ludhiana 141004}
\affiliation{University of Science and Technology of China, Hefei 230026}
\affiliation{Seoul National University, Seoul 151-742}
\affiliation{Soongsil University, Seoul 156-743}
\affiliation{Sungkyunkwan University, Suwon 440-746}
\affiliation{School of Physics, University of Sydney, NSW 2006}
\affiliation{Department of Physics, Faculty of Science, University of Tabuk, Tabuk 71451}
\affiliation{Tata Institute of Fundamental Research, Mumbai 400005}
\affiliation{Excellence Cluster Universe, Technische Universit\"at M\"unchen, 85748 Garching}
\affiliation{Toho University, Funabashi 274-8510}
\affiliation{Tohoku University, Sendai 980-8578}
\affiliation{Department of Physics, University of Tokyo, Tokyo 113-0033}
\affiliation{Tokyo Institute of Technology, Tokyo 152-8550}
\affiliation{Tokyo Metropolitan University, Tokyo 192-0397}
\affiliation{Utkal University, Bhubaneswar 751004}
\affiliation{CNP, Virginia Polytechnic Institute and State University, Blacksburg, Virginia 24061}
\affiliation{Wayne State University, Detroit, Michigan 48202}
\affiliation{Yamagata University, Yamagata 990-8560}
\affiliation{Yonsei University, Seoul 120-749}

  \author{A.~Bala}\affiliation{Panjab University, Chandigarh 160014} 
  \author{V.~Bhardwaj}\affiliation{Nara Women's University, Nara 630-8506} 
  \author{K.~Trabelsi}\affiliation{High Energy Accelerator Research Organization (KEK), Tsukuba 305-0801}\affiliation{The Graduate University for Advanced Studies, Hayama 240-0193} 
  \author{J.~B.~Singh}\affiliation{Panjab University, Chandigarh 160014} 
  \author{A.~Abdesselam}\affiliation{Department of Physics, Faculty of Science, University of Tabuk, Tabuk 71451} 
  \author{I.~Adachi}\affiliation{High Energy Accelerator Research Organization (KEK), Tsukuba 305-0801}\affiliation{The Graduate University for Advanced Studies, Hayama 240-0193} 
  \author{H.~Aihara}\affiliation{Department of Physics, University of Tokyo, Tokyo 113-0033} 
  \author{K.~Arinstein}\affiliation{Budker Institute of Nuclear Physics SB RAS and Novosibirsk State University, Novosibirsk 630090} 
  \author{D.~M.~Asner}\affiliation{Pacific Northwest National Laboratory, Richland, Washington 99352} 
  \author{V.~Aulchenko}\affiliation{Budker Institute of Nuclear Physics SB RAS and Novosibirsk State University, Novosibirsk 630090} 
% ...
\collaboration{The Belle Collaboration}

\begin{abstract}
We report the first observation of $B^0 \to X(3872) (K^{+}\pi^{-})$  and evidence for $B^+ \to X(3872) (K^{0}\pi^{+})$. We measure the product of branching fractions for the former to be ${\cal B}(B^0 \to X(3872) (K^+ \pi^-)) \times {\cal B}(X(3872) \to J/\psi \pi^+ \pi^-) = (7.9 \pm 1.3(\mbox{stat})\pm 0.4(\mbox{syst})) \times 10^{-6}$ and find that $B^{0}\to X(3872) K^{*}(892)^{0}$ does not dominate the $B^{0}\to X(3872)K^{+}\pi^{-}$ decay mode. We also measure ${\cal B}(B^+ \to X(3872) (K^0 \pi^+)) \times {\cal B}(X(3872) \to J/\psi \pi^+ \pi^-) = (10.6 \pm 3.0(\mbox{stat}) \pm 0.9(\mbox{syst})) \times 10^{-6}$. This study is based on the full data sample of 711~fb$^{-1}$ ($772\times 10^6 B\bar B$ pairs) collected at the $\Upsilon(4S)$ resonance with the Belle detector at the KEKB collider.

\end{abstract}


\maketitle

{\renewcommand{\thefootnote}{\fnsymbol{footnote}}}
\setcounter{footnote}{0}

\par 
About a decade ago, the Belle Collaboration discovered 
the $X(3872)$ state~\cite{Choi:2003ue} in the exclusive reconstruction 
of $B^{+} \rightarrow X(3872)(\to J/\psi \pi^+ \pi^-) K^{+}$~\cite{charge_conjugate}. Considerable effort by both experimentalists 
and theorists has been invested to clarify its nature. As a result, we know precisely its mass (3871.69$\pm$0.17)~MeV$/c^2$~\cite{pdg2014}, have 
a stringent limit on its width (less than 1.2~MeV at 90\% confidence 
level)~\cite{Choi:2011prd}
and have a definitive $J^{PC}$ assignment of $1^{++}$~\cite{Aaij:2013zoa}. 
The $X(3872)$ has been observed to decay to several other final states: 
$J/\psi \gamma$~\cite{Bhardwaj:2011prl}, 
$\psi' \gamma$~\cite{Aaij:2014pg}, 
$J/\psi \pi^+ \pi^- \pi^0$~\cite{Sanchez:2010prd} and
$D^{0} {\bar{D}}^{*0}$~\cite{Aushev:2010prd,Aubert:2008prd_babar}. 
The proximity of its mass to the $D^0$-${\bar{D}}^{*0}$ threshold,  along with 
its measured partial decay rates, suggests that it be a loosely bound ``molecule'' 
of $D^0$ and  ${\bar{D}}^{*0}$ mesons~\cite{Swanson:2004pp} or 
an admixture of $D^0 {\bar{D}}^{*0}$ with a charmonium ($c\bar{c}$) 
state~\cite{Swanson:2004pp, Suzuki:2005}. 
Some authors have advanced a QCD-tetraquark interpretation for the $X(3872)$, and predict the existence of charged- and $C$-odd partner states that are nearby in mass~\cite{Maiani:2004vq}.  Experimental searches for charged-~\cite{Choi:2011prd, Aubert:2005prd} and $C$-odd~\cite{Iwashita:2014, Vishal:2013} partners report negative results.  However, since these searches are restricted to states with narrow total widths, the published limits may not apply if the partner states access more decay channels and are thus broader.
More experimental information on the production and decays of the $X(3872)$ 
will shed additional light on its nature. 

\par In this paper, we present the results of searches for $X(3872)$ 
production via the $B^0 \to X(3872) K^+ \pi^-$ and $B^+ \to X(3872) K^0_S \pi^+$ 
decay modes, where the $X(3872)$ decays to 
$J/\psi \pi^+ \pi^-$. The study is based on 711~fb$^{-1}$ of data containing 
$772\times 10^{6}$ $B\bar{B}$ events collected with the Belle 
detector~\cite{Abashian} at the KEKB 
$e^+e^-$ asymmetric-energy collider~\cite{Kurokawa:2003Abe:2103} operating 
at the $\Upsilon(4S)$ resonance. In addition to selecting 
$B \to X(3872) K\pi$ signal events, the same selection criteria isolate 
a rather pure sample of $B \to \psi' K\pi$ events that are used for calibration.

\par The Belle detector is a large solid-angle magnetic
spectrometer that consists of a silicon vertex detector (SVD),
a 50-layer central drift chamber (CDC), an array of
aerogel threshold Cherenkov counters (ACC),
a barrel-like arrangement of time-of-flight
scintillation counters (TOF), and an electromagnetic calorimeter (ECL)
composed of CsI(Tl) crystals. All these detector components are 
located inside a superconducting solenoid coil that provides a 1.5~T
magnetic field.  An iron flux return located outside 
the coil is instrumented to detect $K_L^0$ mesons and to identify
muons (KLM). The detector is described in detail elsewhere~\cite{Abashian}.
\par Monte Carlo (MC) samples are generated for each decay mode using EvtGen~\cite{x_evtgen} and radiative effects are taken into account using the {\scriptsize PHOTOS}~\cite{x_photos} package. The detector response is 
simulated using {\scriptsize GEANT}3~\cite{x_geant}.
\par Charged tracks are required to originate from the interaction 
point (IP). To identify charged kaons and pions, we use a likelihood ratio 
${\cal R}_{K/\pi} = {\cal L}_{K}/({\cal L}_{\pi}+{\cal L}_{K})$, where 
the kaon (pion) likelihood ${\cal L}_{K}$ (${\cal L}_{\pi}$) is calculated using ACC, 
TOF and CDC measurements. For the prompt charged kaon (pion), 
we apply the criterion ${\cal R}_{K/\pi}~({\cal R}_{\pi/K})>0.6$. 
Here, the kaon (pion) identification efficiency is 93\% (95\%) while 
the probability of misidentifying a pion as a kaon (kaon as a pion) 
is 8\% (7\%). For the pion daughters from $\psi'$ or $X(3872)$, we impose 
${\cal R}_{\pi/K}>0.4$; the corresponding pion identification 
efficiency is 99\% and the misidentification probability is 8\%.
Candidates for the $K_S^0\to\pi^+\pi^-$ decay are formed from pairs of 
oppositely charged tracks 
having an invariant mass between 488 and 506~MeV/$c^2$ 
($\pm 4\sigma$ around the nominal mass of $K_{S}^{0}$). 
The $K_S^0$ candidate is also required to satisfy the criteria described 
in Ref.~\cite{Chen:2005dra} to ensure that its decay vertex is 
displaced from the IP. A track is identified as a muon if its
muon likelihood ratio is greater than 0.1, where the muon and hadron likelihoods are calculated by the
track penetration depth and hit scatter in the muon detector (KLM).
An electron track is identified with an electron likelihood greater 
than 0.01, where the electron likelihood is calculated by combining 
$dE/dx$ from the CDC, the ratio of the energy deposited in the ECL 
and the momentum measured by the SVD and the CDC, 
the shower shape in the ECL, ACC information and the position matching between
the shower and the track. With the above selections, the muon (electron) 
identification efficiency is above 90\% and the hadron fake rate is 
less than 4\% (0.5\%).
\par We reconstruct $J/\psi$ mesons in the $\ell^+ \ell^-$ decay channel
($\ell =e~\rm {or}~\mu$) and include bremsstrahlung 
photons that are within 50 mrad of either the $e^+$ or $e^-$ tracks 
[hereinafter denoted as $e^+ e^- (\gamma)$].
The invariant mass of the $J/\psi$ candidate is required 
to satisfy $3.00$~GeV/$c^2$ $< M_{e^+ e^- (\gamma)} <  3.13$~GeV/$c^2$ 
or $3.06$~GeV/$c^2 < M_{\mu^+ \mu^-}  <  3.13$~GeV/$c^2$ (with a distinct lower 
value accounting for the residual bremsstrahlung in the electron mode). 
A mass- and vertex-constrained fit is performed to the selected $J/\psi$ 
candidate to improve its momentum resolution. 
The $J/\psi$ candidate is then combined with a $\pi^+\pi^-$ pair to form an $X(3872)$ ($\psi'$) candidate whose mass must satisfy 
3.82~GeV$/c^2$ $<$ $M_{J/\psi\pi\pi}$ $<$ 3.92~GeV$/c^2$
(3.64~GeV$/c^2$ $<$ $M_{J/\psi\pi\pi}$ $<$ 3.74~GeV$/c^2$).
The dipion mass must also satisfy 
$M_{\pi \pi} > M_{J/\psi\pi\pi} - (m_{J/\psi} + 0.2~{\rm GeV}/c^2)$, where 
$m_{J/\psi}$ is nominal mass. This criterion corresponds to 
$M_{\pi\pi} > 575~(389)$~MeV/$c^2$ for the $X(3872)~(\psi')$ mass region and
reduces significantly the combinatorial background~\cite{Choi:2011prd}
while also flattening the background shape distribution
in $M_{J/\psi \pi \pi}$. 
To suppress the background from $e^+ e^- \to q \bar{q}$ 
(where $q$ = $u$, $d$, $s$, $c$) continuum events, we require $R_2 < 0.4$, 
where $R_2$ is the ratio of the second- to zeroth-order Fox-Wolfram 
moments~\cite{foxwolf}.
\begin{linenomath}
\par To reconstruct a neutral (charged) $B$ meson candidate, a $K^+\pi^-$ 
($K^0_S \pi^+$) candidate is combined with the $X(3872)$ or $\psi'$ 
candidate. We select $B$ candidates using two kinematic variables: the 
energy difference $\Delta E = E_B - E_{\rm beam}$ and the beam-energy 
constrained mass $M_{\rm bc} = (\sqrt{E_{\rm beam}^{2}- p_B^2c^2})/c^2$, where 
${E_{ \rm beam}}$ is the beam energy and $E_B$ and $p_B$ are 
the energy and magnitude of momentum, respectively, of the candidate 
$B$ meson, all calculated in the $e^+ e^-$ center-of-mass (CM) frame.
Only $B$ candidates having $M_{\rm bc} > 5.27$~GeV/$c^2$ and 
$|\Delta E| < 0.1$~GeV are retained for further analysis.
After all selection criteria, approximately 35\% of events have multiple $B$ candidates.
For an event with more than one $B$ candidate, we select the candidate 
having the smallest value of
\begin{equation}
  \chi^2 = \frac{(M_{\rm bc} - 5.2792~\rm{GeV/}{\it{c}}^2)^{2}}{\sigma_{M_{\rm bc}}^{2}} + \frac{\chi^{2}_{B}}{ndf},
\end{equation}
\end{linenomath}
where $\sigma_{M_{\rm bc}}$ is the $M_{\rm bc}$ resolution (estimated to be 2.925~MeV/$c^2$ 
from a fit to $B^0\to\psi'K^+\pi^-$ events), $\chi^{2}_{B}$ is the quality of the vertex fit of all 
charged tracks (excluding the $K_S^0$ daughters), $ndf$ = ($2N-3$) in this fit and $N$ is the number of fitted tracks. The correct candidate is selected in about 60\% of the $B \to X(3872)K\pi$ events with multiple entries.

\par To extract the signal yield of $B\to X(3872)(\to J/\psi\pi^+\pi^-) K\pi$,
 we perform a two-dimensional (2D) unbinned extended maximum 
likelihood fit to the $\Delta E$ and $M_{J/\psi\pi\pi}$ distributions.
For the signal, the $\Delta E$ distribution is parametrized by the sum of a Crystal 
Ball~\cite{crystal_ball} and a Gaussian function while the $M_{J/\psi\pi\pi}$ 
distribution is modeled using the sum of two Gaussians having a common mean.
The 2D probability distribution function (PDF) is a product of the individual 
one-dimensional PDFs, as no sizable correlation is found.
\par The main background contribution in $B \to (J/\psi \pi^+\pi^-) K \pi$ is expected 
to arise from inclusive $B$ decays to $J/\psi$, which is confirmed by the low 
background found in the $J/\psi$ mass sidebands in the data. To study this background,
 we use a large Monte Carlo sample of 
$B \to J/\psi X$ events corresponding to 100 times the integrated luminosity 
of the data sample, and we find that few backgrounds are peaking in the $M_{J/\psi\pi\pi}$ 
distribution (nonpeaking in the $\Delta E$ distribution) and vice versa. The remaining backgrounds are 
combinatorial in nature and are flat in both distributions. This background is parametrized
by first-order Chebyshev polynomial.
\par
\begin{figure}[htbp]
\begin{center}
  \begin{tabular}{cc} 
\includegraphics[width=0.49\textwidth]{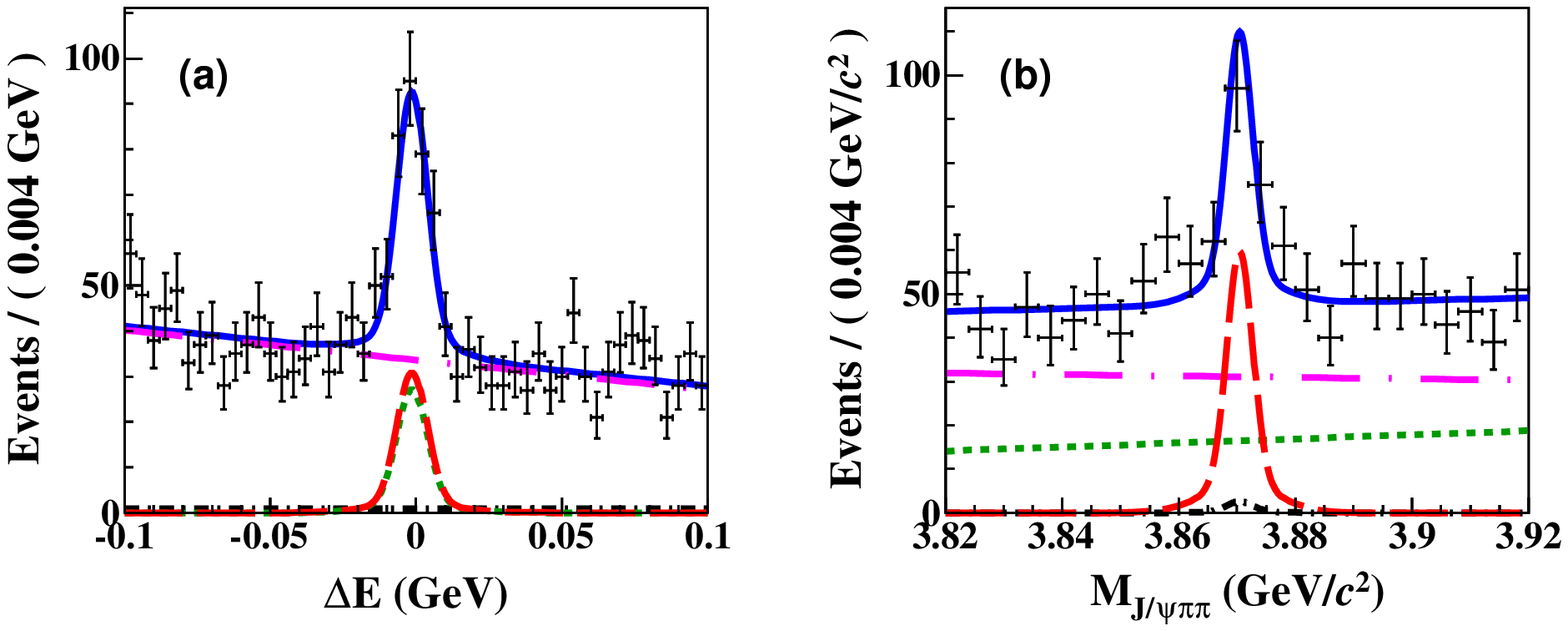}\\
\includegraphics[width=0.49\textwidth]{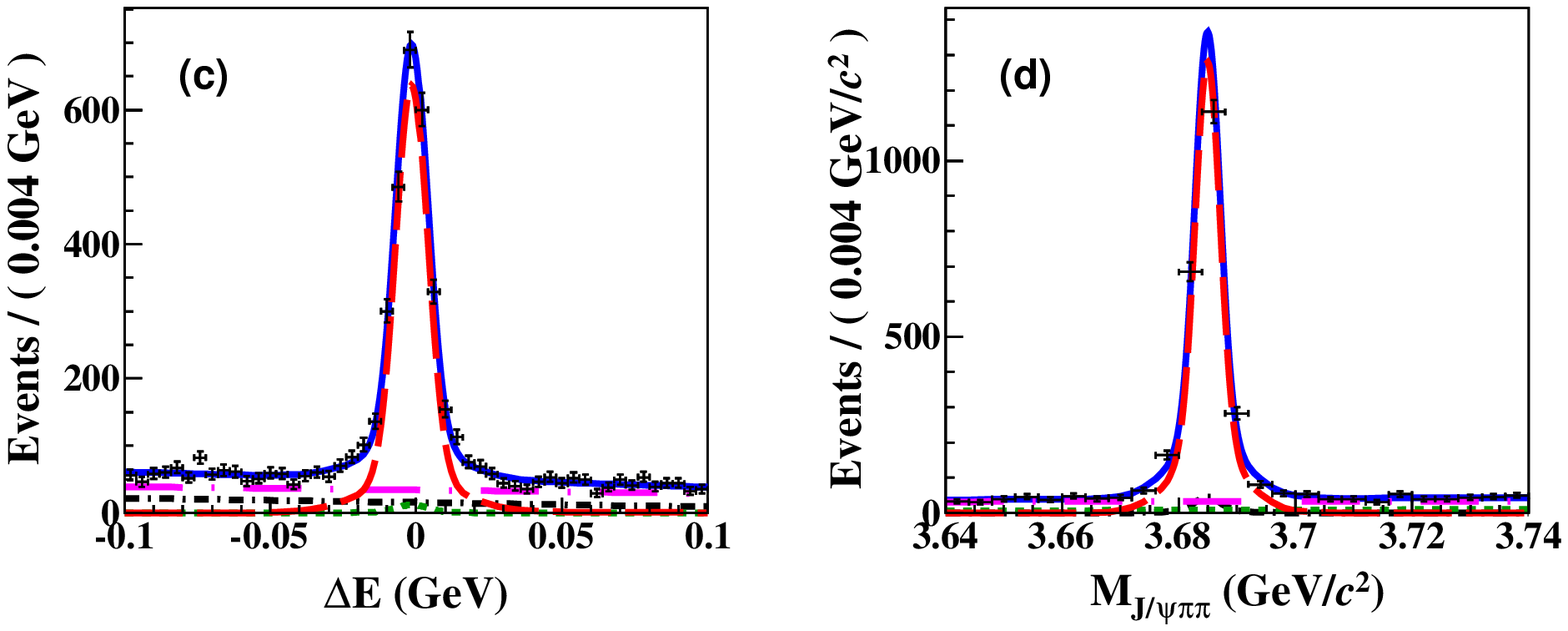}
   \end{tabular}
\caption{Projections of the ($\Delta E$, $M_{J/\psi\pi\pi}$) fit for the 
$B^0\to X(3872)(\to J/\psi\pi^+\pi^-)K^+\pi^-$ decay mode (top) and the 
$B^0\to \psi'(\to J/\psi\pi^+\pi^-)K^+\pi^-$ decay mode (bottom): 
(a) $\Delta E$ distribution for 
3.860~GeV$/c^2< M_{J/\psi\pi\pi} <$ 3.881~GeV$/c^2$,
(b) $M_{J/\psi\pi\pi}$ distribution for $-11$~MeV $< \Delta E <$ 8~MeV,
(c) $\Delta E$ distribution for 3.675~GeV$/c^2< M_{J/\psi\pi\pi} <$ 
3.695~GeV$/c^2$ and (d) $M_{J/\psi\pi\pi}$ distribution for 
$-11$~MeV $< \Delta E <$ 8~MeV.
The curves show the signal (red long-dashed curve)
  and the background components (black dash-dotted line for the component peaking in $M_{J/\psi\pi\pi}$ 
but nonpeaking in $\Delta E$, green dashed line for the one peaking in $\Delta E$ 
but nonpeaking in $M_{J/\psi\pi\pi}$, and magenta long dash-dotted line 
for combinatorial background) as well as the overall fit (blue solid curve).}
\label{fig:signal_enhanced_psi}
\end{center}
\end{figure}

 For the $B^0 \to X(3872) K^+ \pi^-$ decay mode, a 2D fit is performed.  
 The mean and resolution of $M_{J/\psi\pi\pi}$ and $\Delta E$ are fixed 
for the $X(3872)$ region from signal MC samples after being rescaled from the 
results of the $B^{0}\to\psi'K^{+}\pi^{-}$ decay mode. Further, we correct 
the mean of a Gaussian function for the $M_{J/\psi\pi\pi}$ distribution because of a difference in the shift of the 
$\psi'$ and $X(3872)$ reconstructed and generated masses as seen in MC samples. The tail parameters, $\alpha$ and $n$ of the Crystal Ball (CB) function, and the
ratios of the CB's area and width to the corresponding quantities of the
Gaussian component are fixed according to the signal MC simulation.
 The peaking components can be divided into 
two categories: the one peaking in $M_{J/\psi\pi\pi}$ but nonpeaking in $\Delta E$ 
that comes from the $B\to X(3872) X'$ decays where the $X(3872)$ decays in 
$J/\psi\pi^+\pi^-$ (here $X'$ can be any particle), and the other peaking in $\Delta E$ but nonpeaking 
in $M_{J/\psi \pi \pi}$ which comes from a $B$ with the same final state 
where $J/\psi\pi^+\pi^-$ is not from a $X(3872)$. 
The peaking background in $\Delta E$ ($M_{J/\psi \pi \pi}$) is found to have the
same resolution as that of the signal, so the PDF is chosen to be the same 
as the signal PDF, while the nonpeaking background in the other dimension 
is parametrized with a first-order Chebyshev polynomial.
Parameters (slopes) of the background PDFs 
, which are of nonpeaking or combinatorial nature, are allowed 
to vary in the fit. 
The fits are validated on full simulated experiments and no significant bias 
is seen. Figure~\ref{fig:signal_enhanced_psi}~(top)
shows the signal-enhanced projection plots for the $B^0\to X(3872) (K^+ \pi^-)$ decay mode. 
The result of the fit and branching fractions derived are listed in 
Table~\ref{table:tab_results}.
We find a clear signal for $B^{0}\to X(3872)K^{+}\pi^{-}$  of $116 \pm 19$, 
signal events corresponding to a significance~\cite{cousinhighland} (including systematic uncertainties related to the signal yield as mentioned in Table~\ref{table:tab_br_syst}) of 7.0 standard deviations 
($\sigma$), and measure the product of branching fractions to be
$\mathcal{B}(B^{0}\to X(3872)K^{+}\pi^{-}) \times \mathcal{B}(X(3872)\to J/\psi\pi^{+}\pi^{-}) = (7.9 \pm 1.3 (\mbox{stat}) \pm 0.4 (\mbox{syst}))\times 10^{-6}$.
The efficiency used for estimating the branching fraction is calculated from the individual efficiencies and the 
fractions of the different components obtained in the $(K^+ \pi^-)$ mass, as 
explained below. The statistical significance is estimated using the value 
of $\sqrt{-2\ln(\mathcal{L}_{0}/\mathcal{L}_{\rm max})}$ where 
$\mathcal{L_{\rm max}}$ ($\mathcal{L_{\rm 0}}$) denotes the likelihood 
value when the yield is allowed to vary (fixed to zero). 

\begin{figure}
\begin{center}
  \begin{tabular}{c}  
\includegraphics[width=0.24\textwidth]{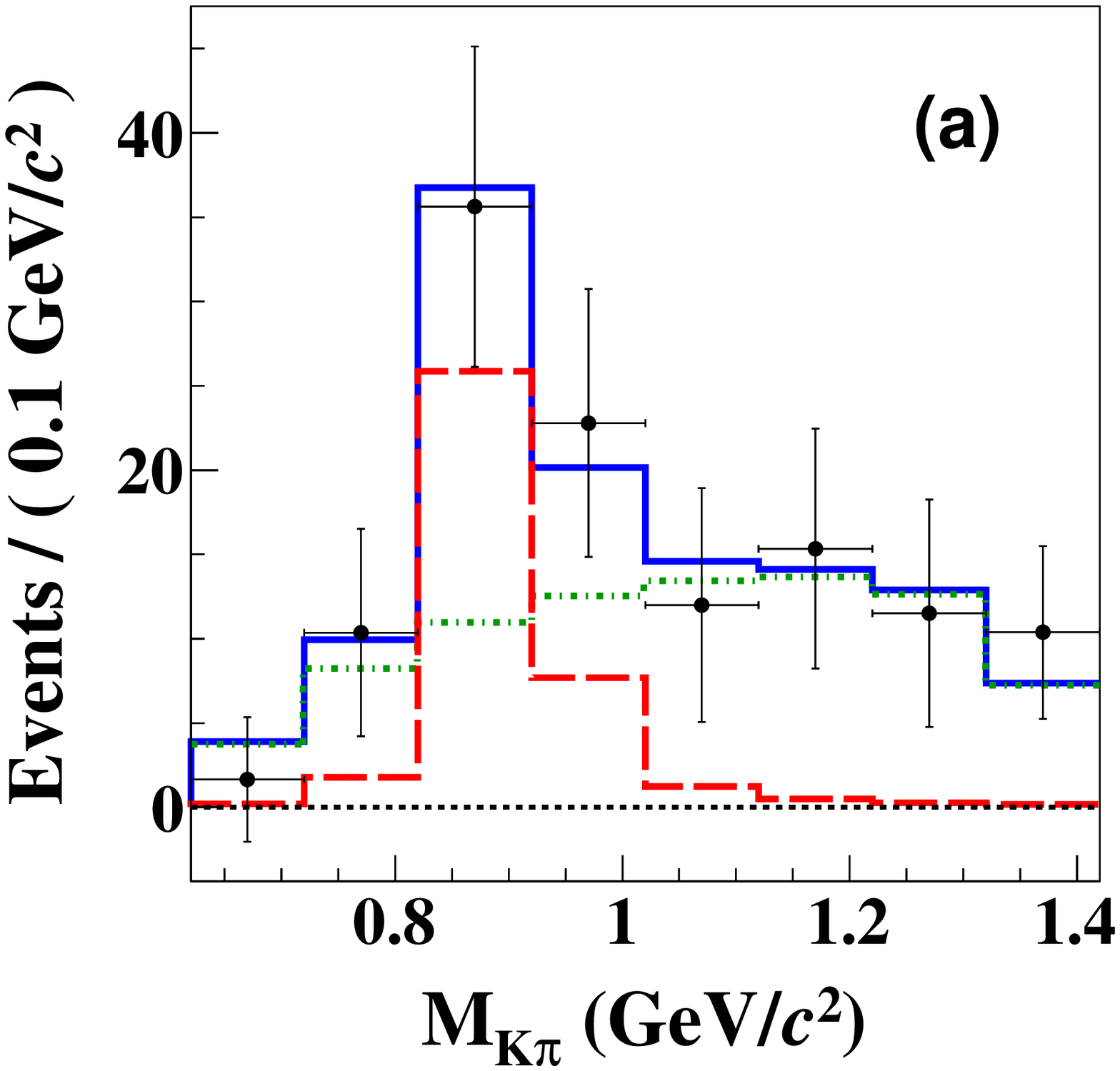}
\includegraphics[width=0.24\textwidth]{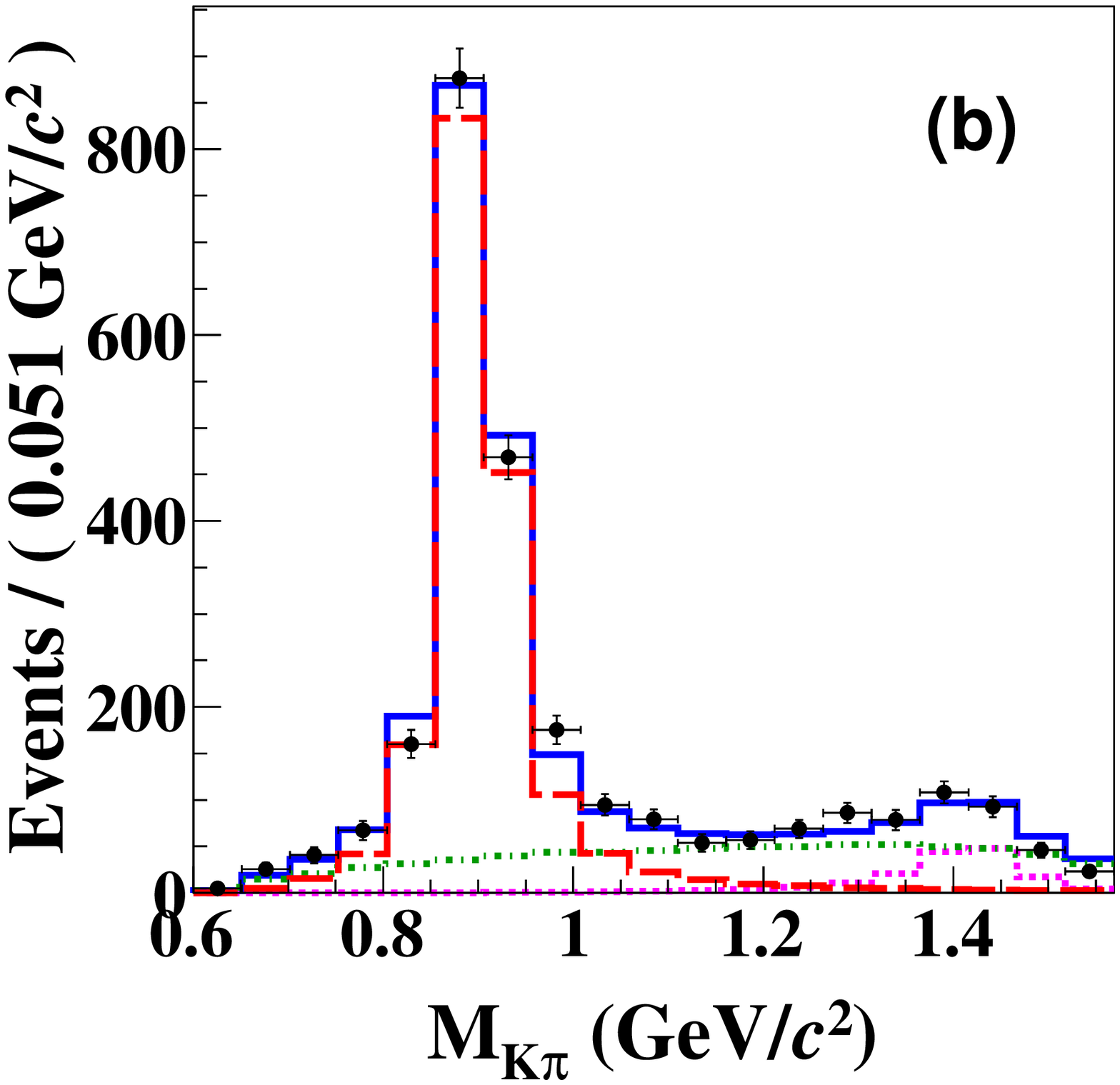}
  \end{tabular}
  \caption{Fit to the background-subtracted $M_{K\pi}$ distribution: 
(a) for the $B^0\to X(3872) (K^+\pi^-)$ decay mode, the curves show the
    $B^0\to X(3872) K^*(892)^0$ (red long-dashed lines), 
    $B^0\to X(3872) (K^+\pi^-)_{\rm NR}$ (green dot-dashed lines),
    as well as the overall fit (blue solid lines).
    (b) for the $B^0\to \psi' (K^+\pi^-)$ decay mode, the curves show the 
    $B^0\to \psi' K^*(892)^0$ (red long-dashed lines), 
    $B^0\to \psi' (K^+\pi^-)_{\rm NR}$ (green dot-dashed lines),
    $B^0\to \psi' K_{2}^*(1430)^{0}$ (magenta dashed lines)
    as well as the overall fit (blue solid lines). 
    }
  
\label{fig:data_binned_psi}
\end{center}
\end{figure}

The above fit is validated on the calibration mode $B^0\to\psi'K^+\pi^-$. In contrast to the $X(3872)$ mass region, the mean and resolution in both distributions 
($M_{J/\psi\pi\pi}$ and $\Delta E$) are allowed to vary in the fit. 
Figure~\ref{fig:signal_enhanced_psi} (bottom) shows the signal-enhanced 
projection plots for the $B^0\to \psi' (K^+ \pi^-)$ decay mode. 
We measure the branching fraction to be 
$\mathcal{B}(B^{0} \to \psi'K^{+}\pi^{-})$ =
$(5.79 \pm 0.14(\mbox{stat}))\times 10^{-4}$,
consistent with an independent Belle result based on a Dalitz-plot 
analysis~\cite{Chilikin:2013tch}.
%
\begin{table*}[!hbpt]
\caption{Signal yield (Y) from the fit, weighted efficiency ($\epsilon$) after particle-identification 
correction, significance ($\Sigma$) and measured $\mathcal{B}$ for 
$B^0\to X(3872)K^+\pi^-$ and $B^+\to X(3872)K^0\pi^+$.
The first (second) uncertainty represents a statistical 
(systematic) contribution.}
\begin{center}
\begin{tabular}{lcccc}
\hline
\hline
Decay mode & Y $\;$ & $\;$ $\epsilon$ (\%) &
$\;$ $\Sigma$ ($\sigma$) &
 ${\cal B}{(B \to X(3872) K \pi)} \times$ ${\cal B}(X(3872) \to J/\psi \pi^+ \pi^-)$ \\
\hline
$B^0 \to X(3872) K^+\pi^-$ & $116 \pm 19$ & 15.99 & 7.0 & $(7.9 \pm 1.3 \pm 0.4)\times 10^{-6}$ \\
$B^+ \to X(3872) K^{0}\pi^{+}$ & $35 \pm 10$ & 10.31 & 3.7 & $(10.6 \pm 3.0 \pm 0.9) \times 10^{-6}$ \\
\hline
\hline
\end{tabular}
\end{center}
\label{table:tab_results}
\end{table*}
\begin{table}[!htbp]
\caption{Summary of the systematic uncertainties in percent.}
\begin{center}
\begin{tabular}{lccc}
\hline
\hline
Source & $X(3872) $ & $X(3872)$  \\
 & $K^+\pi^-$ & $K^{0}\pi^{+}$   \\
\hline
Lepton ID              & 3.4 &  3.4  \\
Kaon ID                & 1.1 &  ...   \\
Pion ID                & 2.5 &  3.2   \\
PDF modeling                 &  $^{+1.8}_{-1.3}$ & $^{+4.2}_{-4.9}$\\
Tracking efficiency          & 2.1 &  2.5 \\
$K^0_S$ reconstruction   &  ...  &  0.7  \\ 
$N_{B\bar{B}}$    & 1.4 & 1.4 \\
Secondary ${\cal B}$   & 0.4 & 0.4  \\
Efficiency & $0.6$ & $1.0$ \\
Fit bias               & 0.6 &  3.1  \\
\hline
Total & $5.4$  &  $8.0$ \\
\hline
\hline
\end{tabular}
\end{center}
\label{table:tab_br_syst}
\end{table}
\begin{linenomath}
\par Further, to determine the contribution of the $K^*(892)$ and other intermediate states,
we perform a 2D fit to $\Delta E$ and $M_{J/\psi\pi\pi}$ in each bin of $M_{K\pi}$ 
(0.1 GeV/$c^2$ wide bins of $M_{K\pi}$ in the range $[0.62, 1.42]$ GeV/$c^2$), 
which provides a background-subtracted $M_{K \pi}$ signal distribution.
All parameters of the signal PDFs for $M_{J/\psi\pi\pi}$ and $\Delta E$ 
distributions are fixed from the previous 2D fit to all events. 
We perform a $\chi^2$ fit to  the $M_{K\pi}$  distribution using 
$K^*(892)^0$ and $(K^{+}\pi^{-})_{\rm NR}$ components, which are histogram PDFs
obtained from MC samples. Note that the 
$B^{0}\to X(3872) {K_{2}}^{*}(1430)^{0}$ decay is kinematically suppressed. 
We do not consider the interference between the $K^*(892)$ 
and nonresonant component since the number of candidates is not large 
enough to make a full amplitude analysis. 
The resulting fit result is shown in Figure~\ref{fig:data_binned_psi}(a).
We obtain $38 \pm 14$ ($82 \pm 21$) signal events for the
$B^{0}\to X(3872) K^{*}(892)^{0}$ ($B^{0}\to X(3872) (K^+\pi^-)_{\rm NR}$) decay 
mode, whose sum is consistent with the total yield obtained from the global fit. 
This corresponds to a 3.0$\sigma$ significance (including 
systematic uncertainties related to the signal yield) for the
$B^{0}\to X(3872)(\to J/\psi \pi^{+} \pi^{-}) K^{*}(892)^0$ decay mode, 
and a product of branching fractions of 
${\cal B}(B^{0} \to X(3872)K^*(892)^0) \times {\cal B}(X(3872) \to J/\psi \pi^+ \pi^-)  = (4.0 \pm 1.5(\mbox{stat}) \pm 0.3(\mbox{syst})) \times 10^{-6}$.
The ratio of branching fractions is
\begin{equation}
\label{eqn:fraction}
\begin{split}
\frac{\mathcal{B}(B^0 \to X(3872)K^{*}(892)^{0})\times \mathcal{B}(K^{*}(892)^{0} \to K^{+}\pi^{-})}{\mathcal{B}(B^0 \to X(3872) K^+\pi^- )} \\
 = 0.34 \pm 0.09(\mbox{stat}) \pm 0.02(\mbox{syst}).
\end{split}
\end{equation}
In the above ratio, all systematic uncertainties cancel except 
those from the PDF model, fit bias and efficiency variation over the Dalitz distribution.
We generate pseudoexperiments to estimate the significance of the $\chi^2$ fit.
\end{linenomath}
\par The $B^0 \to \psi' K^+ \pi^-$ mode is analyzed with the same procedure, with
0.051 GeV/$c^2$ wide bins, due to the copious yield, and in the $M_{K\pi}$ range 
[0.600, 1.569]~GeV$/c^2$. We perform a $\chi^2$ fit to the obtained $M_{K\pi}$ 
signal distribution again to extract the contributions of the $K\pi$ 
nonresonant and resonant components. For this purpose, we use 
histogram PDFs obtained from MC samples of several possible components of the $(K^+\pi^-)$ system: ${K}^{*}(892)^{0}$, 
$K_2^*(1430)^{0}$ and nonresonant $K^+\pi^-$ $((K^{+}\pi^{-})_{\rm NR})$; in the last case, $B^0\to\psi'(K^+\pi^-)_{\rm NR}$  
is generated uniformly in phase space.
The fit result is shown in Figure~\ref{fig:data_binned_psi}(b).
The $K^*(892)$ dominates and we measure 
$\mathcal {B}(B^{0}\to\psi'{K}^{*}(892)^{0}) = 
(5.88 \pm 0.18 (\mbox{stat})) \times 10^{-4}$,  
which is consistent with the world average~\cite{pdg2014}. 
\par In contrast to $B^{0}\to \psi' (K^+\pi^-)$ [where the ratio 
of branching fractions is $0.68 \pm 0.01(\mbox{stat})$], 
$B^{0}\to X(3872) K^{*}(892)^{0}$ does not dominate in
the $B^{0}\to X(3872)K^{+}\pi^{-}$. 
%
\par
\begin{figure}[!htpb]
\begin{center}
  \begin{tabular}{cc}  
\includegraphics[width=0.49\textwidth]{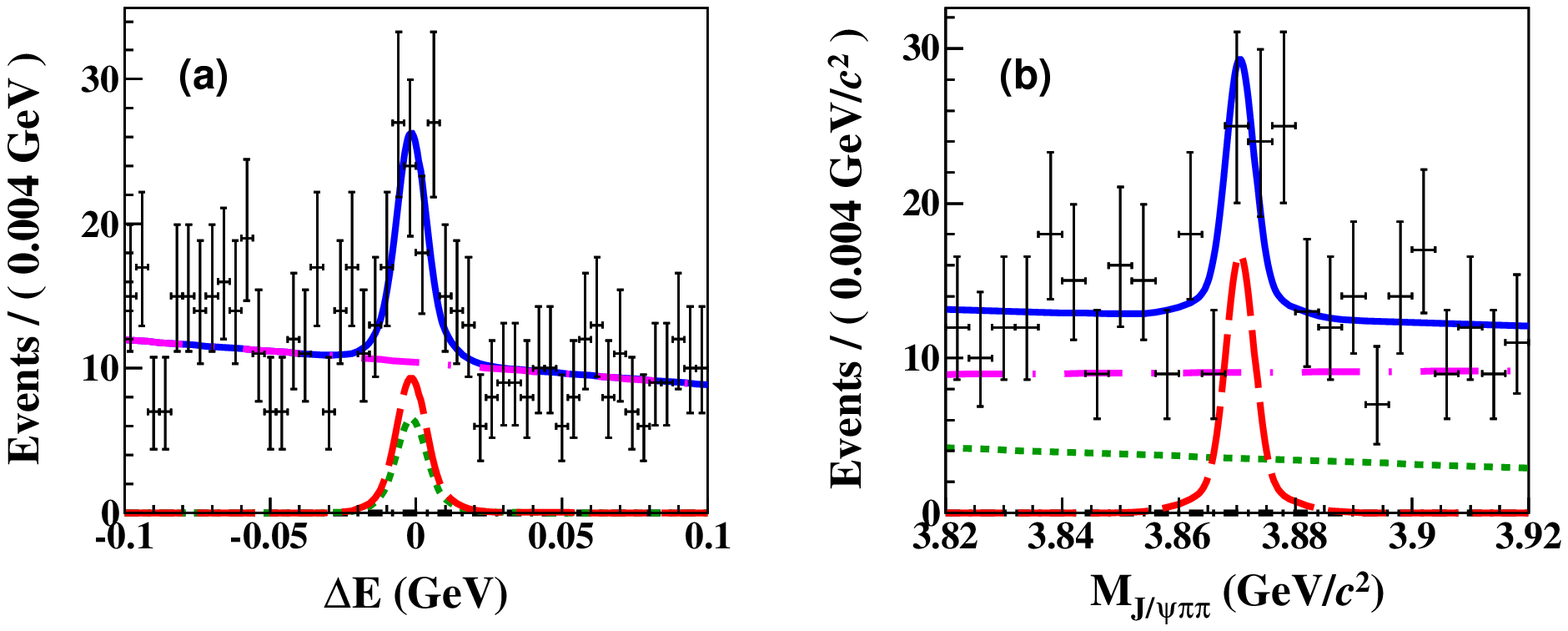}\\
\includegraphics[width=0.49\textwidth]{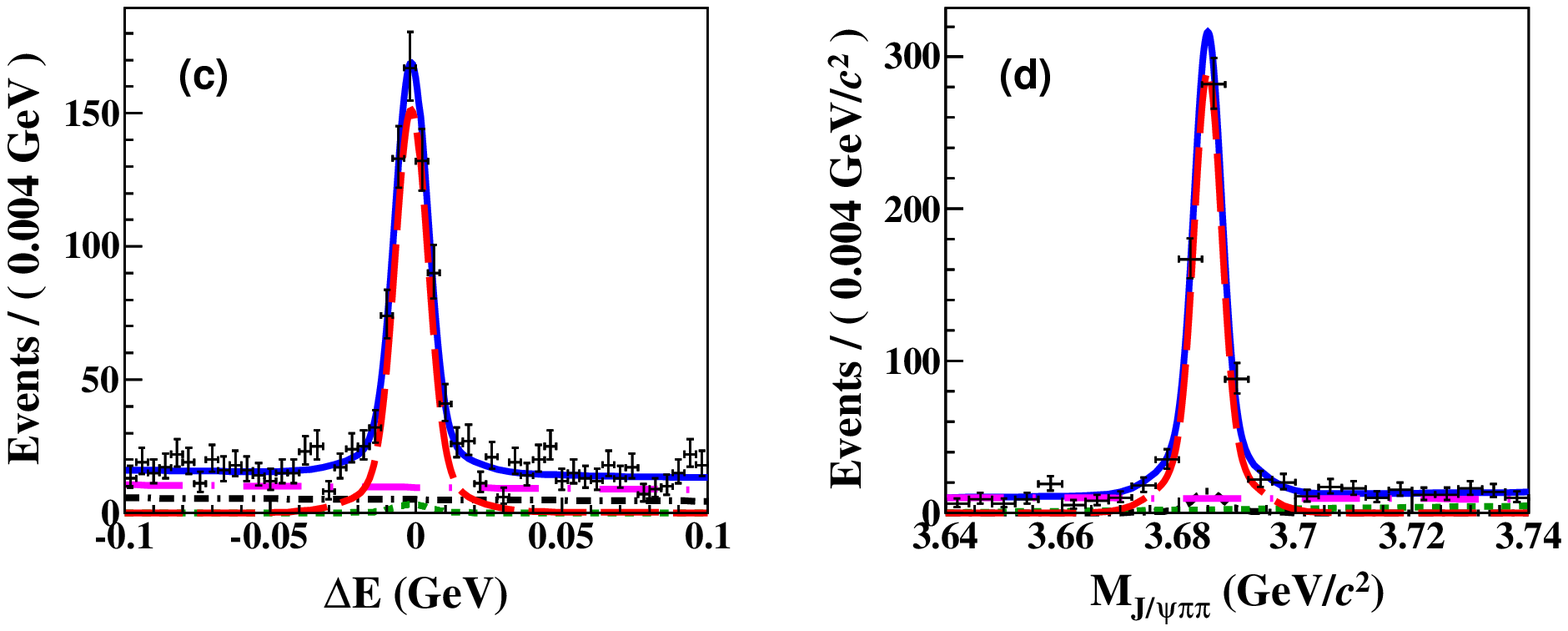} 
     \end{tabular}
\caption{Projections of the ($\Delta E$, $M_{J/\psi\pi\pi}$) fit 
for the $B^\pm \to X(3872)(\to J/\psi\pi^+\pi^-)K_S^{0}\pi^{\pm}$ decay mode (top) 
and for the $B^{\pm}\to \psi'(\to J/\psi\pi^{+}\pi^{-})K_S^{0}\pi^{\pm}$ decay 
mode (bottom):
  (a) $\Delta E$ distribution for 3.859~GeV$/c^2< M_{J/\psi\pi\pi} <$ 
3.882~GeV$/c^2$,
  (b) $M_{J/\psi\pi\pi}$ distribution for $-11$~MeV $< \Delta E <$ 9~MeV,
  (c) $\Delta E$ distribution for 3.675 GeV$/c^2< M_{J/\psi\pi\pi} <$ 
3.695~GeV$/c^2$ and
  (d) $M_{J/\psi\pi\pi}$ distribution for $-11$~MeV $< \Delta E <$ 9~MeV.
  The curves show the signal (red long-dashed curves)
  and the background components (black dash-dotted curves for the component peaking in $M_{J/\psi\pi\pi}$ 
but nonpeaking in $\Delta E$, green dashed lines for the one peaking in $\Delta E$ 
but nonpeaking in $M_{J/\psi\pi\pi}$, and magenta long dash-dotted lines 
for combinatorial background) as well as the overall fit (blue solid lines).}
\label{fig:2d_data_ks}
\end{center}
\end{figure}

 We also investigate the decays 
$B^{+}\to X(3872)(\to J/\psi \pi^{+} \pi^{-})(K^{0}\pi^{+})$.
The PDFs of $\Delta E$ and $M_{J/\psi\pi\pi}$ are the same as those for 
the neutral $B$ meson decay mode.
The projections of the 2D fit for $B^{+}\to X(3872)(\to J/\psi \pi^{+} \pi^{-}) (K^{0}\pi^{+})$ in the
signal-enhanced regions are shown in Figs.~\ref{fig:2d_data_ks}(a) and ~\ref{fig:2d_data_ks}(b). 
We find $35 \pm 10$ events for the
$B^{+}\to X(3872)(\to J/\psi \pi^{+} \pi^{-}) (K^{0}\pi^{+})$ decay mode, 
corresponding to a 3.7$\sigma$ significance (including 
systematic uncertainties). The product of branching fractions is 
${\cal B}(B^{+} \to X(3872)K^0\pi^{+}) \times {\cal B}(X(3872) \to J/\psi \pi^+ \pi^-)  = (10.6 \pm 3.0(\mbox{stat}) \pm 0.9(\mbox{syst})) \times 10^{-6}$.
The above fit is validated for the $\psi'$ mass region.
The projections of the 2D fit for $B^{+}\to \psi'(\to J/\psi \pi^{+} \pi^{-}) (K^{0}\pi^{+})$ in the
signal-enhanced regions are shown in Figs.~\ref{fig:2d_data_ks}(c) and ~\ref{fig:2d_data_ks}(d). 
The branching fraction for 
$B^{+}\to \psi'(\to J/\psi \pi^{+} \pi^{-}) (K^{0}\pi^{+})$ 
is $(6.00 \pm 0.28 (\mbox{stat})) \times 10^{-4}$, while the 
world average of this quantity is $(5.88 \pm 0.34)\times 10^{-4}$. 
%
\begin{table}[!htbp]   
\caption{Summary of systematic uncertainties (in percent) used for the 
$M_{K\pi}$ background-subtracted fit in $B^0\to X(3872) K^+ \pi^-$.}
\begin{center}
\begin{tabular}{lcc}
\hline
\hline
Source & $X(3872) K^*(892)^0$   \\
\hline
Lepton ID              & 3.4   \\
Kaon ID                & 1.1  \\
Pion ID                & 2.6    \\
PDF modeling                  &  $^{+1.5}_{-1.4}$  \\
Tracking efficiency          & 2.1  \\
$N_{B\bar{B}}$    & 1.4  \\
Secondary ${\cal B}$   & 0.4  \\
MC statistics          & 0.2  \\
Fit bias               &  4.6   \\
\hline
Total     & $7.0$ \\
\hline
\hline
\end{tabular}
\end{center}
\label{table:summ_sys_binned}
\end{table}
\par Equal production of neutral and charged 
$B$ meson pairs in the $\Upsilon (4S)$ decay is assumed. Secondary 
branching fractions used for calculation of $\mathcal{B}$ are taken 
from Ref.~\cite{pdg2014}. 
Systematic uncertainties are summarized in Tables~\ref{table:tab_br_syst} 
and~\ref{table:summ_sys_binned}.
A correction for small differences in the signal detection efficiency 
between signal MC events and data due to lepton, kaon and pion 
identification differences is applied; samples of $J/\psi \to \ell^+\ell^-$ and 
$D^{*+} \to D^0(\to K^-\pi^+) \pi^+$ decays are used to estimate this correction.
 The uncertainties on these corrections are included in the 
systematic error. The uncertainty due to the fitting 
model is obtained by varying all fixed parameters by $\pm{1}\sigma$ and 
considering the corresponding change in the yield as the systematic error. 
The uncertainties due to tracking efficiency, $K_S^0$ reconstruction  
and $N_{B\bar{B}}$ are estimated to be 0.35\% per track, 0.7\%  
and 1.4\%, respectively. 
The systematic uncertainty due to secondary branching fractions is included. 
The systematic uncertainty for the efficiency arises from the 
limited MC statistics and the efficiency variation over the Dalitz distribution is also considered.
Small biases in the fitting procedure, estimated in the ensemble study, 
are also considered as a source of systematic uncertainty. For this study we perform a fit to
100 pseudoexperiments after embedding signal events obtained from MC samples to each inclusive MC sample.
All the above stated systematic uncertainties are added in quadrature and result in a total systematic uncertainty of 5.4\%, 8.0\%, 7.0\% for $B^0\to X(3872) K^+ \pi^-$, $B^+\to X(3872) K_S^0 \pi^+$ 
and $B^0\to X(3872) K^*(892)^0$, respectively.
\par In summary, we report the first observation of the $X(3872)$ in the decay 
$B^0 \to X(3872)K^+\pi^-$, $X(3872) \to J/\psi \pi^+ \pi^-$. 
The result for the $X(3872)$, where $B^0 \to X(3872) K^*(892)^0$
does not dominate the  $B^0 \to X(3872) (K^+\pi^-)$ decay, is in marked
contrast to the $\psi'$ case.
We have checked for a structure in 
the $X(3872)\pi$ and $X(3872)K$ invariant masses and found no evident peaks.
We measure ${\cal B}(B^0 \to X(3872) (K^+ \pi^-)) 
\times {\cal B}(X(3872) \to J/\psi \pi^+ \pi^-) = 
(7.9 \pm 1.3(\mbox{stat}) \pm 0.4(\mbox{syst}))\times 10^{-6}$
and ${\cal B}(B^+ \to X(3872)K^{0}\pi^{+}) 
\times {\cal B}(X(3872) \to J/\psi \pi^+ \pi^-) = (10.6 \pm 3.0(\mbox{stat}) \pm 0.9(\mbox{syst})) \times 10^{-6}$. \\

\par We thank the KEKB group for the excellent operation of the
accelerator; the KEK cryogenics group for the efficient
operation of the solenoid; and the KEK computer group,
the National Institute of Informatics, and the 
PNNL/EMSL computing group for valuable computing
and SINET4 network support.  We acknowledge support from
the Ministry of Education, Culture, Sports, Science, and
Technology (MEXT) of Japan, the Japan Society for the 
Promotion of Science (JSPS), and the Tau-Lepton Physics 
Research Center of Nagoya University; 
the Australian Research Council and the Australian 
Department of Industry, Innovation, Science and Research;
Austrian Science Fund under Grant No.~P 22742-N16 and P 26794-N20;
the National Natural Science Foundation of China under Contracts 
No.~10575109, No.~10775142, No.~10875115, No.~11175187, and  No.~11475187; 
the Ministry of Education, Youth and Sports of the Czech
Republic under Contract No.~LG14034;
the Carl Zeiss Foundation, the Deutsche Forschungsgemeinschaft
and the VolkswagenStiftung;
the Department of Science and Technology of India; 
the Istituto Nazionale di Fisica Nucleare of Italy; 
National Research Foundation (NRF) of Korea Grants
No.~2011-0029457, No.~2012-0008143, No.~2012R1A1A2008330, 
No.~2013R1A1A3007772, No.~2014R1A2A2A01005286, No.~2014R1A2A2A01002734, 
No.~2014R1A1A2006456;
the Basic Research Lab program under NRF Grant No.~KRF-2011-0020333, 
No.~KRF-2011-0021196, Center for Korean J-PARC Users, No.~NRF-2013K1A3A7A06056592; 
the Brain Korea 21-Plus program and the Global Science Experimental Data 
Hub Center of the Korea Institute of Science and Technology Information;
the Polish Ministry of Science and Higher Education and 
the National Science Center;
the Ministry of Education and Science of the Russian Federation and
the Russian Foundation for Basic Research.
the Slovenian Research Agency;
the Basque Foundation for Science (IKERBASQUE) and 
the Euskal Herriko Unibertsitatea (UPV/EHU) under program UFI 11/55 (Spain);
the Swiss National Science Foundation; the National Science Council
and the Ministry of Education of Taiwan; and the U.S.\
Department of Energy and the National Science Foundation.
This work is supported by a Grant-in-Aid from MEXT for 
Science Research in a Priority Area (``New Development of 
Flavor Physics'') and from JSPS for Creative Scientific 
Research (``Evolution of Tau-lepton Physics'').


\begin{thebibliography}{99}
\bibitem{charge_conjugate}
Charge-conjugate decays are included unless explicitly
stated otherwise.
\end{thebibliography}
\end{document}